\documentclass{article}
\usepackage[sorting=ynt]{biblatex} 
\usepackage{caption}
\usepackage{subcaption}
\usepackage{authblk}
\usepackage{amsmath}
\usepackage{graphicx}
\usepackage{amssymb}

\begin{document}

\title{Sound generation mechanisms in a collapsible tube}

\author[1,2]{Marco Laudato\footnote{laudato@kth.se. The following article has been submitted to The Journal of Acoustical Society of America. After it is published, it will be found at https://pubs.aip.org/asa/jasa.}}
\author[2]{Elias Zea}
\author[1]{Elias Sundstr\"om}
\author[2]{Susann Boij}
\author[1]{Mihai Mihaescu}			
\affil[1]{FLOW Research Center, Department of Engineering Mechanics, KTH Royal Institute of Technology, Stockholm, SE-10044, Sweden.}
\affil[2]{The Marcus Wallenberg Laboratory for Sound and Vibration Research, Department of Engineering Mechanics, KTH Royal Institute of Technology, Stockholm, SE-10044, Sweden}




\maketitle

\begin{abstract}
Collapsible tubes can be employed to study the sound generation mechanism in the human respiratory system. The goals of this work are (a) to determine the airflow characteristics connected to three different collapse states of a physiological tube and (b) to find a relation between the sound power radiated by the tube and its collapse state. The methodology is based on the implementation of computational fluid dynamics simulation on experimentally validated geometries. The flow is characterized by a radical change of behavior before and after the contact of the lumen. The maximum of the sound power radiated corresponds to the post-buckling configuration. The idea of an acoustic tube law is proposed. The presented results are relevant to the study of self-excited oscillations and wheezing sounds in the lungs.
\end{abstract}

\section{Introduction}
\label{sec:intro}

In daily medical practice, auscultation is among the most cost-effective and timely diagnostic tools for diseases of the respiratory and cardiovascular system~\cite{bohadana2014fundamentals}. Different pathologies are characterized by specific sounds (sometimes called \textit{adventitious sounds}) that can guide the physicians toward identifying and assessing the disease. From an acoustic perspective, the problem of sound generation in the human body and its propagation represents a challenging question that is far from being completely understood~\cite{palnitkar2020sound,hassan2019system}.

From a biomechanical perspective, for respiratory and cardiovascular pathologies, the generation of adventitious sounds is often correlated to the interaction between the biological fluid (air or blood) and the human vessel (airways, or veins and arteries)~\cite{iccer2014classification}. The interest in quantitatively modeling the Fluid-Structure Interaction (FSI) phenomenology and the corresponding sound generation mechanisms in the context of different diseases is justified by the ambition to develop and improve sound-based diagnostic tools. In this perspective, thanks to more accessible large-scale high performance computational resources, Computational Fluid Dynamics (CFD) represents a powerful tool able to furnish a 4D description of mass transportation in the human body and the corresponding FSI dynamics~\cite{sikkandar2019computational}.

The current computational power allows to study cardiovascular~\cite{itatani2022hemodynamic} and respiratory~\cite{schoder2020hybrid, lasota2023anisotropic} diseases in both patient-specific~\cite{mylavarapu2013planning, kraxberger2023alignment} and idealized geometries~\cite{shabbir2023impact, li2023time}. One remarkable example of an idealized geometry model of a human vessel is a \textit{collapsible tube}. Collapsible tubes can reproduce all the relevant physical behavior of a human vessel and are widely employed in both numerical models~\cite{babilio2023static} and clinical practice~\cite{wellman2014test}. The advantage of idealized geometries for studying pathophysiological flows is not only found in the lesser computational cost. Numerical models based on CFD simulation with the ambition to provide a reliable quantitative analysis need to be experimentally validated. The possibility to access the large amount of independent high-quality experimental data~\cite{gregory2021experimental, bertram2003experimental, zarandi2021effect} targeting collapsible tubes under different conditions makes the validation step possible and reliable. Moreover, by reducing the complexity of a biological system to its very fundamental parts, it is possible to identify the main physical mechanisms driving its dynamics. For these reasons, collapsible tubes play a major role in analyzing many different pathologies of the cardiovascular and respiratory systems.

One example of application of collapsible tubes in biomedical acoustics is the study of respiratory wheezes. Wheezing is one of the most common adventitious sounds generated in the lungs. It is connected to several different pathologies of human lungs like asthma~\cite{gern2010abcs} and chronic obstructive pulmonary disease~\cite{barnes2009asthma}. Tonal features characterize it and can happen during both inhalation and expiration. The sound generation mechanism of wheezing sounds is associated with the onset of self-excited oscillations in the airways due to the interaction between air and the conduct~\cite{heil2010self}. It is possible to study such FSI-induced oscillations by implementing numerical~\cite{whittaker2010predicting} and experimental~\cite{orucc2007experimental} studies involving collapsible tubes. However, the connection between the FSI and the corresponding generated acoustic signal is still to be completely understood~\cite{zhang2023fully}. 

Another medical research field where collapsible tubes are widely employed is the study of the collapse of the upper airways in pathologies like Obstructive Sleep Apnoea (OSA). OSA patients suffer from recurrent pharynx collapses during sleep, resulting in apnoea and consequent neurological disturbances~\cite{dubessy2023sleepiness}. The diagnosis and assessment of the OSA severity is performed through an expensive procedure called polysomnography. The main goal is to estimate the level of collapsibility of the pharynx in terms of the so-called critical pressure, i.e., the value of the intramural pressure corresponding to the collapse of a compliant tube~\cite{kazemeini2022critical}. Collapsible tubes are also widely employed in the study of the circulatory system~\cite{siviglia2013steady}. One interesting research line is connected to pulse wave imaging ultrasound in stenotic arteries~\cite{mobadersany2023pulse}, in which the carotid is modeled using a collapsible tube undergoing a pulse displacement. Other research directions include the analysis of stenotic arteries~\cite{kumar2022dynamics}, venous collapse prevention~\cite{liu2022development}, the coupling between cardiac cycle and coronary perfusion~\cite{munneke2022closed}, and the study of choroidal blood flow~\cite{spaide2020choroidal}. 

Several studies in the acoustics literature have used collapsible tube models for applications to voice generation. Collapsible tubes have been used to model the vocal cords using an electrical analogy~\cite{Conrad1980}. Since the tube behaves as a flow-controlled nonlinear resistance for steady flows, similar to a transistor, Conrad tested whether vocal cords have the same property. Titze proposed a physical model for small-amplitude oscillations in the vocal folds based on the body-cover hypothesis~\cite{Titze1988}, which is supported by experimental observations in excised larynges---a physical model of the vocal-fold mucosa, and in human subjects~\cite{Chan1999}. In the context of laryngeal cancer, Tourinho \textit{et al}. have used a collapsible tube model for tracheoesophageal (TE) phonation~\cite{Tourinho2021}, a common speech rehabilitation method. The results showed that the tube model reproduced the fundamental frequency and spectral characteristics of TE speech. Recently, Li \textit{et al}. trained a neural network to predict the glottal flow waveform~\cite{Li2021}, using the collapsible tube model from Cancelli and Pedley~\cite{Cancelli1985}. Li's model was validated using experimental data from rabbits with excised larynx. 

In most of the studies in the literature, collapsible tubes have been employed to study biological fluid flows, self-excited oscillations, collapse critical pressure, or to produce acoustic reduced-order models. However, the nature of the fluid-borne sound generation mechanisms in collapsible tubes relevant for biomedical applications and their relation with fluid mechanics structures is still an open question. This work addresses these problems by studying the fluid behavior in three distinguished collapse phases: pre-buckling, post-buckling, and post-contact. The methodology is based on CFD and Computational Aeroacoustics (CAA) modeling. The two goals of this work are: 1) to determine the main fluid characteristics related to sound generation under different collapse conditions, and 2) to relate the overall sound power level generated by the flow with the collapsible state of the tube. This last point is also motivated by a similar experimental investigation on urethral obstruction modeled as a collapsible tube~\cite{terio1991acoustic}.

The structure of the paper is as follows. Sec.~\ref{sec:2} introduces the phenomenology of a collapsible tube under physiological conditions. In Sec.~\ref{sec:numModel}, the details of the numerical model implemented in this work are discussed. Sec.~\ref{sec:soundGen} concerns the sound generation mechanisms due to the interaction of the fluid flow with the geometries corresponding to three different stages of the collapse. In Sec.~\ref{sec:sourceChar}, the analysis of the acoustic power radiated by the collapsible tube under the three different collapse phases is discussed. The onset of acoustic waves in the tube domain is investigated in Sec.~\ref{sec:onsetwaves}. Finally, discussion and conclusions can be found in Sec.~\ref{sec:conclusions}.

\section{Collapsible tube}
\label{sec:2}
\begin{figure}[h!]
    \centering
    \includegraphics[width=0.5\textwidth]{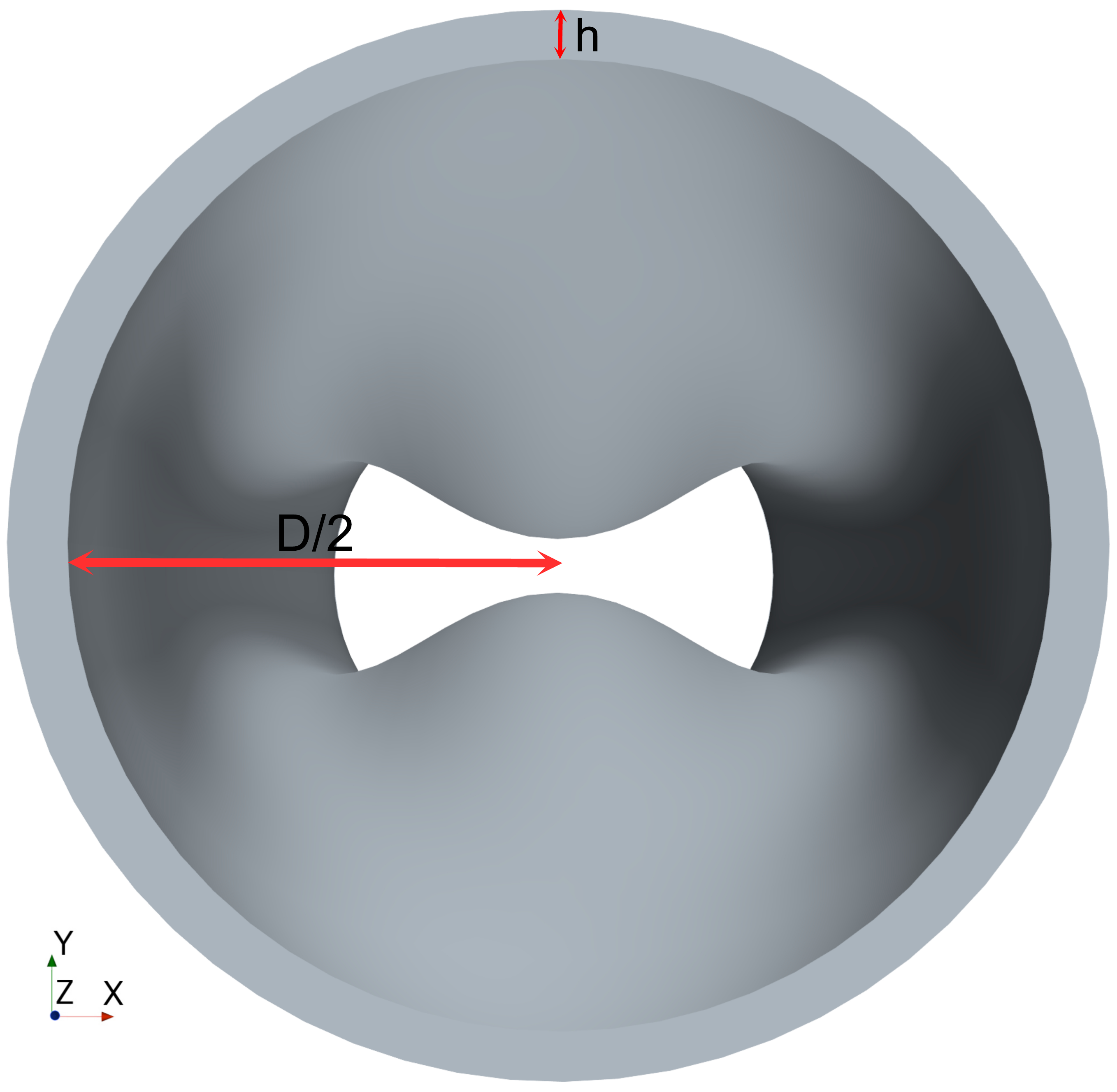}\\
    \includegraphics[width=0.5\textwidth]{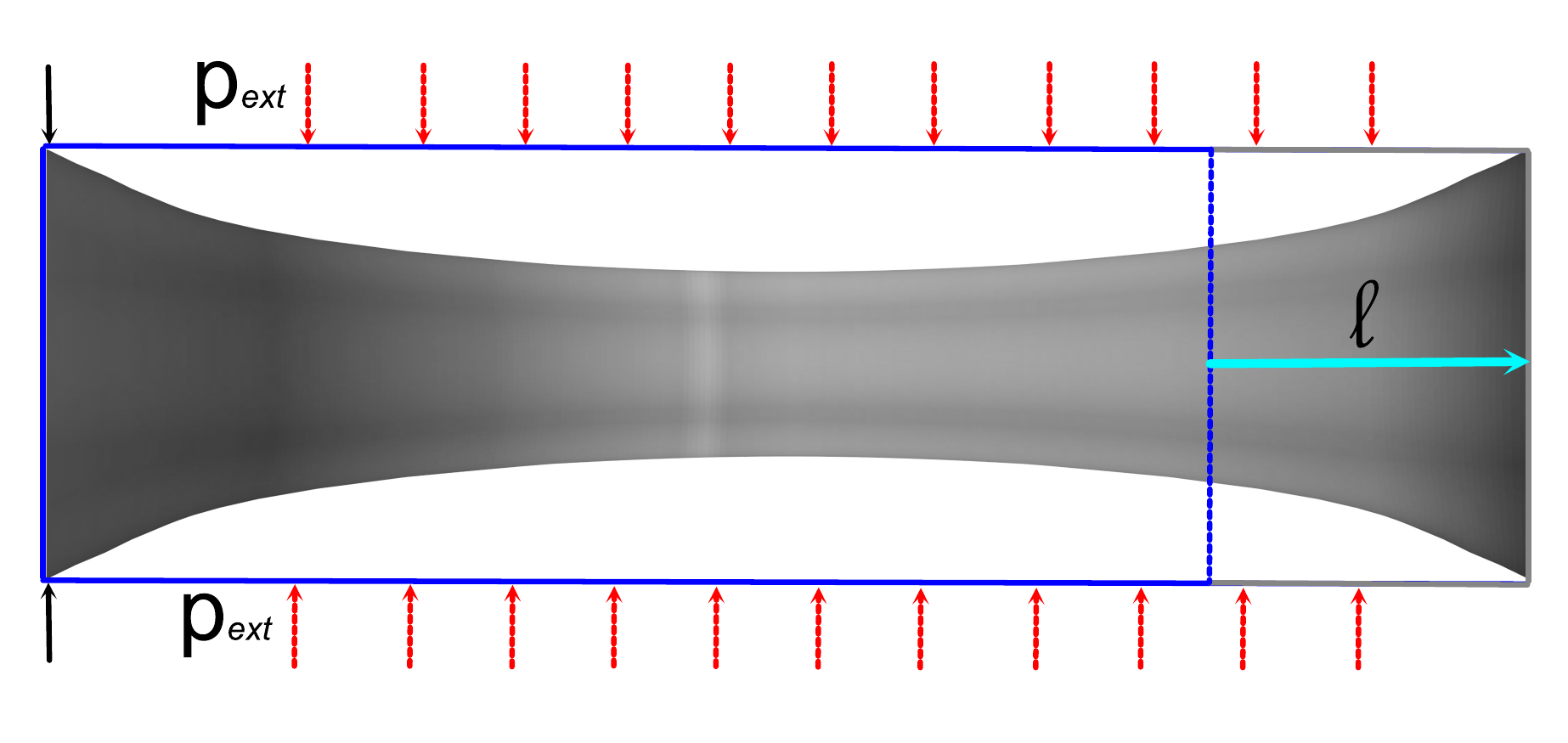}
    \caption{(Color online) Top panel: radial cross-sectional view of the domain pictured in its post-buckling configuration. The (half) diameter $D/2$ and the thickness $h$ are sketched. Bottom panel: lateral view of the actual configuration of the domain under the action of the boundary conditions $p_{ext}$ (red dashed arrows) and $l$ (light blues arrow). The two black solid lines indicate the clamped boundary. The blue box corresponds to the initial configuration of the system.}
    \label{fig:BCs}
\end{figure}

The geometry of a collapsible tube is described by three non-dimensional parameters (see Fig.~\ref{fig:BCs}). The length-to-diameter ratio $d=l_0/D$, the thickness-to-diameter ratio $\gamma=h/D$, and the axial pre-stretch ratio $l=L/l_0$. 

\noindent
The values of these parameters have been chosen to be relevant for biomedical applications~\cite{horsfield1968morphology, hoppin1977axial} and are listed in Tab.~\ref{tab:geom_param}.

In this study, one short side of the tube is clamped while the other is stretched by an amount indicated by the pre-stretch parameter $l$ and then clamped. Under these assumptions, the collapse is only driven by the so-called \textit{intramural pressure} i.e., the pressure difference between the interior and the exterior of the tube:

\begin{equation}
    p_{intr}=p_{int}-p_{ext}\,.
\end{equation}

\begin{figure}[h!]
    \centering
    \includegraphics[width=0.82\textwidth]{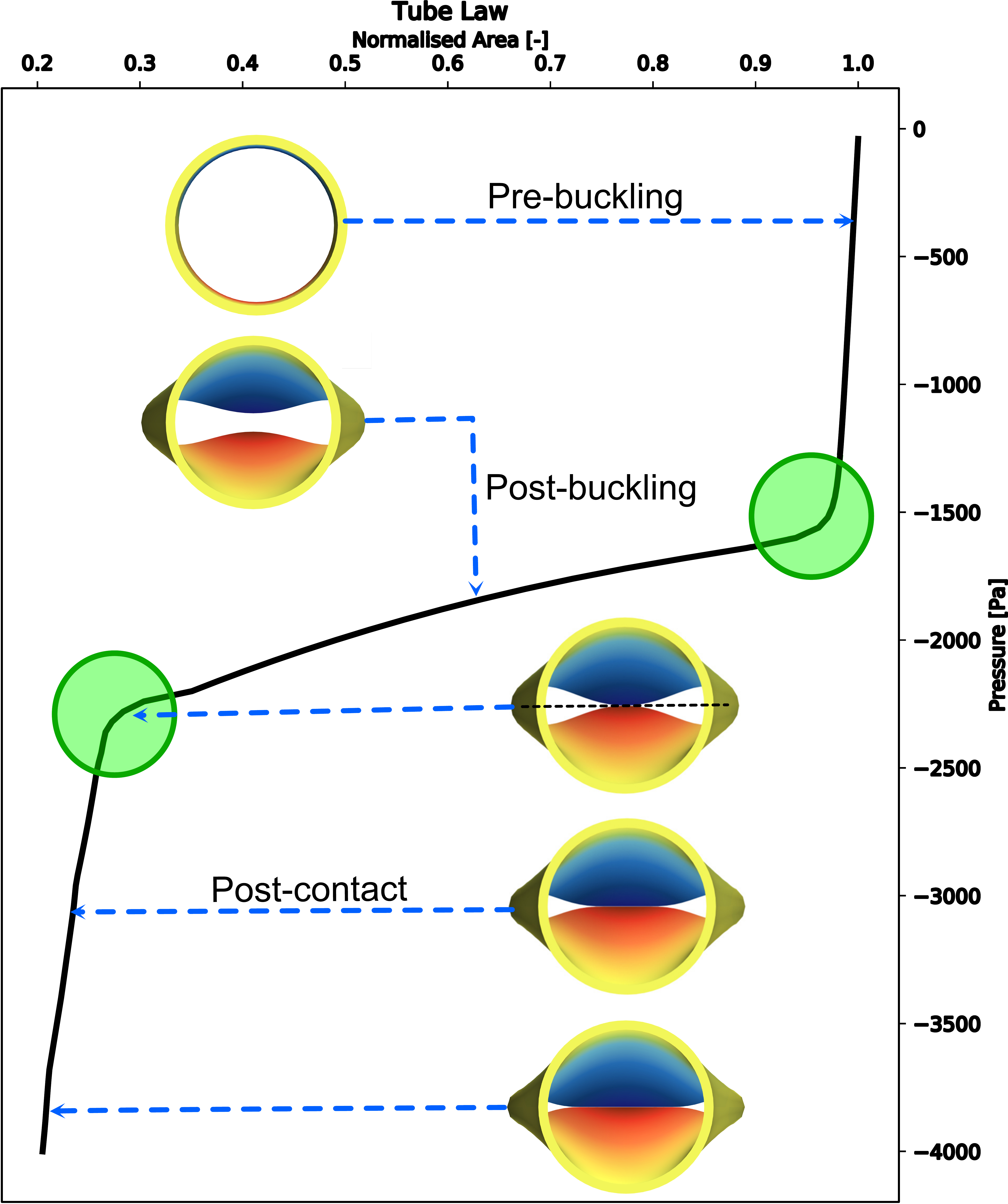}
    \caption{(Color online) The tube law. The right axis indicates the intramural pressure, while the top axis indicates the area of the central cross-section normalised on the initial area. The green areas represents the regions of the transition from the pre-buckling to the post-buckling phase and to the post-contact configuration.}
    \label{fig:tubelaw}
\end{figure}

\noindent
When this difference is negative, the tube starts to collapse. The collapse is described by the so-called \textit{tube law}~\cite{whittaker2010rational}, which relates the area of the central cross-section of the tube and the intramural pressure (see Fig.~\ref{fig:tubelaw}). The tube collapses for small negative values of the intramural pressure by maintaining its axisymmetric shape (\textit{pre-buckling} phase). When the intramural pressure assumes larger negative values, the tube undergoes a buckling phenomenon, resulting in a non-axisymmetric shape (\textit{post-buckling} phase). Under most physiological conditions, the tube's cross-section is characterized by a 2-lobe shape~\cite{han2013artery}. Finally, for even larger negative values of the intramural pressure, the internal walls of the tube touch each other (\textit{post-contact phase})~\cite{flaherty1972post}. At first, the contact region is one point, and it becomes a line for more negative intramural pressures (see Fig.~\ref{fig:tubelaw}).

The transition between the three phases of the collapse (pre-buckling, post-buckling, and post-contact) is marked by two particular values of the intramural pressure called \textit{buckling critical pressures} and \textit{contact critical pressure}. The value of such critical pressures depends on both geometric and elastic parameters of the system~\cite{kozlovsky2014general, zarandi2021effect}. In~\cite{laudato2023buckling}, the authors have shown that the buckling critical pressure can be estimated by treating the buckling as a second-order phase transition (see also~\cite{turzi2020landau} for a rigorous proof for a 1D ring). In~\cite{laudato2023analysis}, a treatment for estimating the contact critical pressure is derived.

The fluid behavior in a collapsible tube under different collapse scenarios has been extensively studied from theoretical~\cite{10.1115/1.3426281}, numerical~\cite{heil1997stokes}, and experimental~\cite{gregory2021experimental} points of view. However, understanding the relation between the tube's characteristic flow structures and the corresponding acoustic features remains an open query. Some relevant studies have been conducted in~\cite{alenius2015large, abom2006aero}, where the authors have studied the aeroacoustics of an orifice plate in a duct using the two-port scattering method. This method, however, targets the characterization of the scattering matrix of the system for an already existing acoustic signal~\cite{boden1995modelling}. Moreover, the system's geometry under analysis does not reflect the 3D behavior of a human vessel.

The present work lies in the larger project of establishing a relation between the acoustics of a human vessel and its collapse level, with the potential development of new diagnostic tools for respiratory and cardiovascular diseases. The following section presents the numerical model implemented in this study. The goal is to perform an in-depth analysis of the fluid behavior under the three collapse phases (pre-buckling, post-buckling, and post-contact) of a collapsible tube. The flow features responsible for the onset of acoustic waves propagating in the tube are studied. Finally, the relation between the acoustic power level generated in the system and the tube's collapse level is presented, supporting the existence of an \textit{acoustic tube law}.

\begin{table}[h!]
    \centering
    \begin{tabular}{|c|c|c|}
        \hline
        $d$ [-] & $\gamma$ [-]& $l$ [-]\\
        \hline
         $3$ $(2.5-6)$&$0.06$ $(0.03-0.1)$&$1.25$ $(1-1.6)$\\
         \hline
    \end{tabular}
    \caption{Values of the geometric parameters of the system. The values in parenthesis correspond to the physiological range in the human airways.}
    \label{tab:geom_param}
\end{table}

\section{Numerical model}
\label{sec:numModel}
It is possible to study the airflow under the three collapse states of a physiological tube and the corresponding acoustic field by employing numerical methods. The numerical model is implemented in the commercial software Siemens Star-CCM+ (v. 2210). The simulation strategy consists of two steps: 

\begin{enumerate}
    \item Solid-only simulation of a collapsible tube under the effect of an isotropic external pressure. The goal is to compute the resulting deformed configurations of the tube corresponding to the pre-buckling, post-buckling, and post-contact phases of the collapse. These will be compared for validation with experimental results and used as fluid domain geometries in the second step.
    \item Simulations of the velocity and pressure field of air flowing in the three fixed collapsed configurations obtained in the previous step. Since air is treated as a compressible ideal gas, the resulting acoustic pressure fluctuations are predicted at the same time.
\end{enumerate}

\subsection{Solid elastic model}
\begin{figure}[h!]
    \centering
    \includegraphics[width=\textwidth]{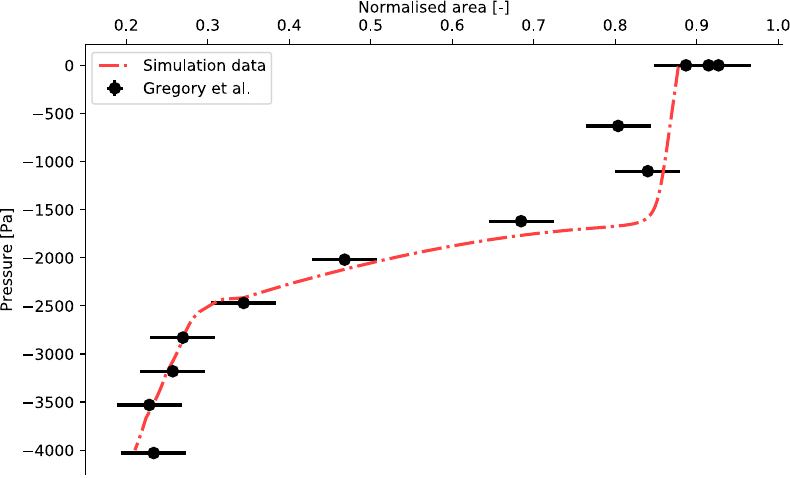}
    \caption{(Color online) Comparison between the tube law obtained via the presented numerical model (red dashed line) versus the corresponding experimental data from \cite{gregory2021experimental}.}
    \label{fig:exp_val}
\end{figure}

The reference configuration geometry is a 3D cylindrical flexible tube with finite thickness and it is implemented in the built-in CAD software in Star-CCM+. The values of the geometric parameters $(d,\gamma, l)$ are listed in Tab.~\ref{tab:geom_param} and are typical for human airways. The system is subject to the following boundary conditions (see Fig.~\ref{fig:BCs}). One short side of the tube is clamped whereas the other one is first stretched by a quantity related to the parameter $l$ and then clamped. The external walls of the domain are under the effect of an isotropic inward pressure $p_{ext}$ which linearly increases in time. Due to the absence of flow, the internal pressure vanishes, $p_{int}=0$, and the intramural pressure depends only on the external pressure, $p_{intr}=-p_{ext}(t)$. The following relation describes the time dependence of the external pressure:

\begin{equation}
\label{eq:p_ext}
    p_{ext}(t) = \frac{P}{\tau}t\,,
\end{equation}

\noindent
where $P$ is the maximum external pressure and $t\in[0,\tau]$ are the time steps. By changing the value of $P$, it is possible to obtain, at time $t=\tau$, the desired collapse state of the tube. If $P<P_{cr}^{b}$, i.e., smaller than the buckling critical pressure, the tube will be in its pre-buckling phase. Analogously, if $P_{cr}^{b}<P<P_{cr}^c$ or $P_{cr}^c<P$, the resulting collapse state will correspond to the post-buckling and post-contact phase, respectively, where $P_{cr}^c$ is the contact critical pressure. The values corresponding to the three different collapse states are $1000$~Pa, $2000$~Pa, and $3000$~Pa for the pre-buckling, post-buckling, and post-contact configuration, respectively (see also Fig.~\ref{fig:exp_val}) and are compatible with typical physiological conditions of the airways. Indeed, under normal breathing conditions, the intramural pressure is approximately $2000$~Pa~\cite{lai1991pleural}. However, for asthma patients, for which the collapse of the airways is often manifested, these values can reach up to $10000$~Pa~\cite{evans2009assessment}, depending on the size of the patient~\cite{lausted2006maximum}. For values of $p_{ext}>P_{cr}^c$, the internal walls of the tube (the \textit{lumen}) touch each other. The contact is handled employing a repulsive virtual plane~\cite{laudato2023analysis}, which prevents the intersection of the solid domain. A sensitivity study of the ratio $P/\tau$ has been performed in~\cite{laudato2023buckling}, and it is not reported here for brevity. For all the following analyses, $\tau=2.5$~s is used. 

Under such boundary conditions, the system undergoes large deformation that cannot be adequately treated with linear elasticity~\cite{laudato2023buckling}. Therefore, the tube is modeled as a Neo-Hookean material whose strain energy potential is defined as

\begin{equation}
\label{eq:strain_en}
    \mathfrak{U} = \Psi(I-3)+\frac{\Phi}{2}(\Tilde{I}-1)^2\,,
\end{equation}

\noindent
where $I$ and $\Tilde{I}$ are the first and second invariants of the right Cauchy-Green tensor, respectively. The coefficients $\Psi$ and $\Phi$ are defined as

\begin{equation}
\label{eq:strain_coeff}
    \Psi=\frac{E}{4(1+\nu)},\qquad\Phi=\frac{E\nu}{(1-2\nu)(1+\nu)}\,,
\end{equation}

\noindent
where $E$ is the Young's modulus and $\nu$ is the Poisson ratio. The values of these elastic parameters are the same as the samples used in~\cite{gregory2021experimental}, whose experimental data will be used to validate the numerical results. 

At each time step ($\Delta t=0.1$~s), the value of the intramural pressure and the area of the central cross-section are registered. It is then possible to determine the corresponding tube law. The validation of the numerical model has been performed by comparison with a public experimental data set~\cite{gregory2017collapse, gregory2021experimental} obtained via 3D camera measurements of the area of the central cross-section of the tube. A digital replica of the experimental specimen has been implemented in Star-CCM+, and the value of the maximum pressure $P$ has been set to match the experimental conditions. The comparison between the numerical and experimental tube laws shows a fair match (see Fig.~\ref{fig:exp_val}).

\begin{figure}[h!]
    \centering
    \begin{subfigure}[b]{0.29\textwidth}
        \includegraphics[width=\textwidth]{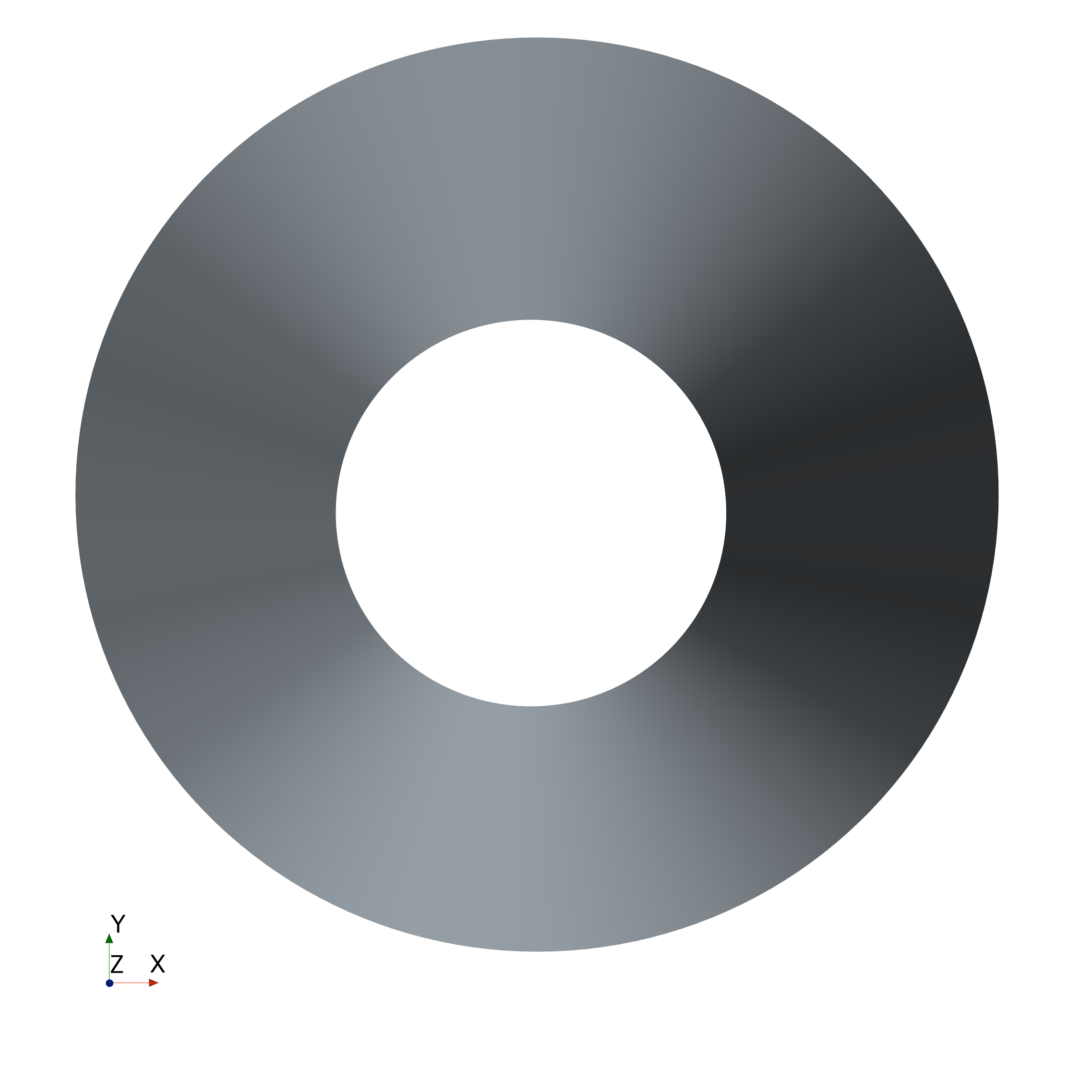}
    \end{subfigure}
    \begin{subfigure}[b]{0.29\textwidth}
        \includegraphics[width=\textwidth]{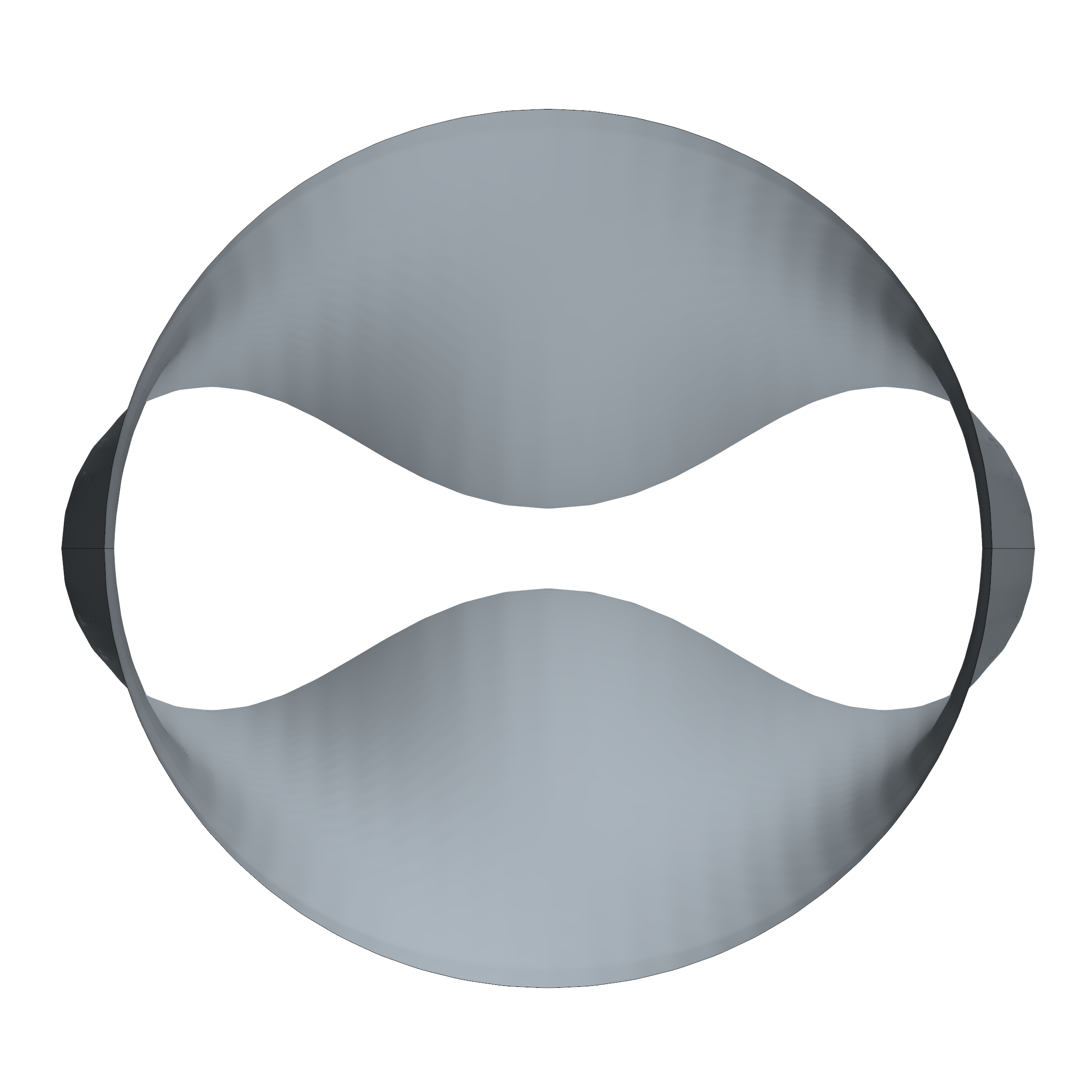}
    \end{subfigure}
    \begin{subfigure}[b]{0.29\textwidth}
        \includegraphics[width=\textwidth]{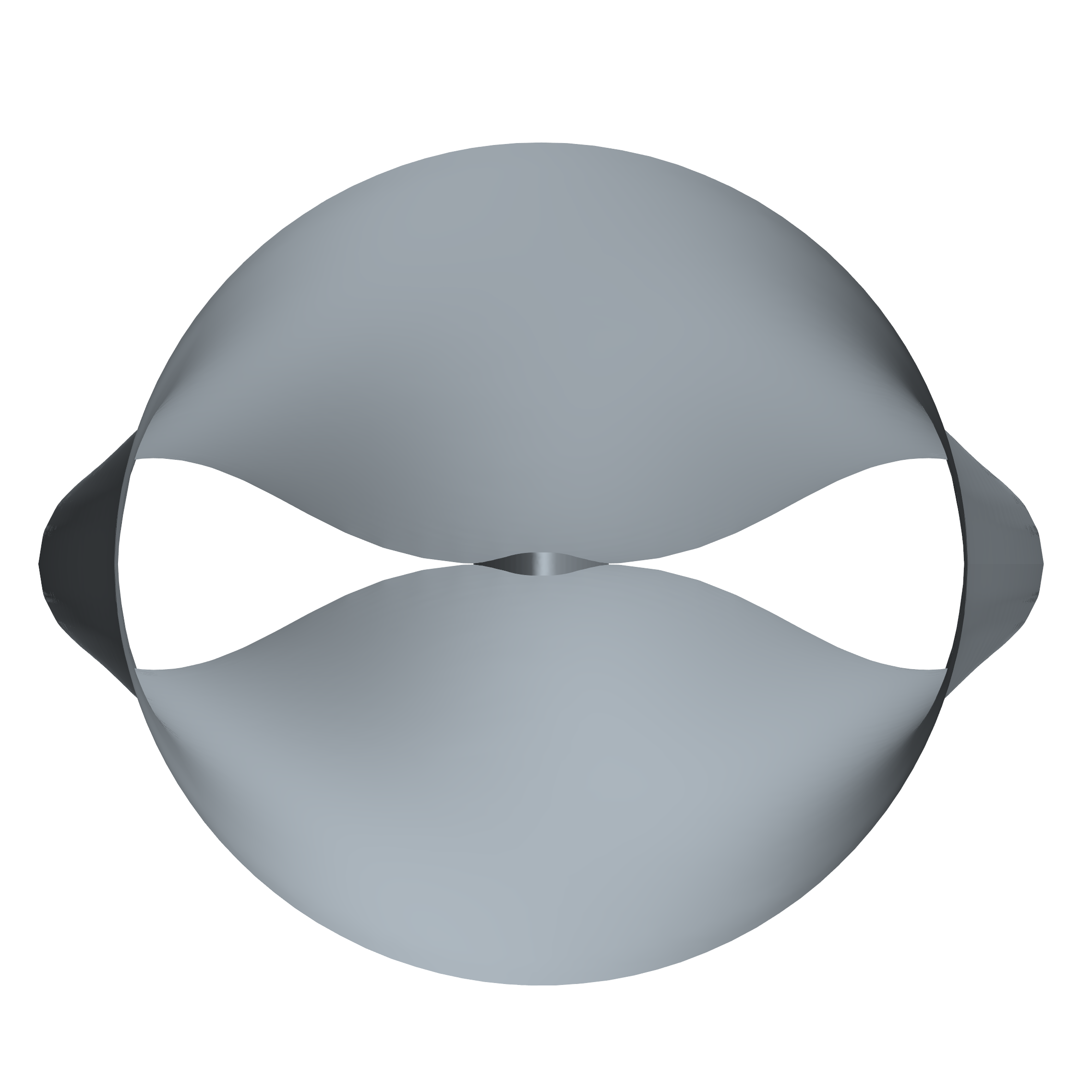}
    \end{subfigure}
    \caption{From left to right, the cross-sections of the fluid domains obtained from the solid model simulations for the pre-buckling, post-buckling, and post-contact configurations.}
    \label{fig:cross-sections}
\end{figure}

The output of these solid-only simulations consists of three experimentally validated deformed configurations of a collapsible tube, corresponding to the pre-buckling, post-buckling, and post-contact phase (see Fig.~\ref{fig:cross-sections}). Such deformed configurations will be employed as numerical domains for the subsequent fluid flow and acoustic field simulations.

\subsection{Fluid model}
\begin{figure*}
    \centering
    \includegraphics[width=\textwidth]{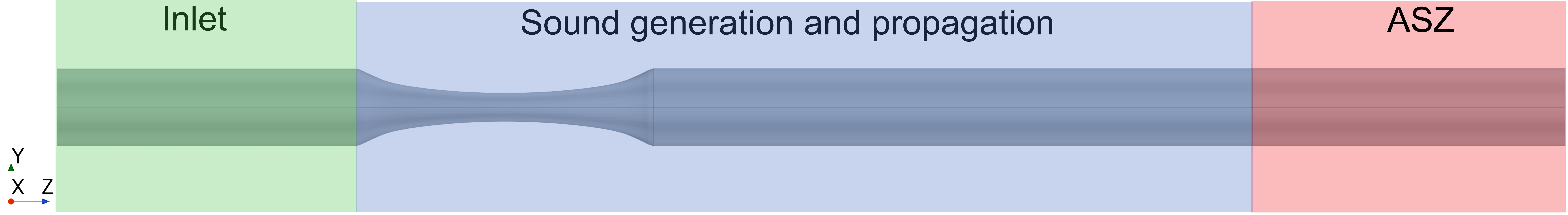}
    \caption{(Color online) Lateral view of the computational domain of the fluid numerical model. In green is highlighted the inlet region. The blue rectangle shows the region of interest and represents the sound generation and propagation region. Finally, the acoustic suppression zone is highlighted in red.}
    \label{fig:fluidDomain}
\end{figure*}

The three geometrical computational domains depicted in Fig.~\ref{fig:cross-sections} are used to simulate the air velocity and pressure fields. The domain is extruded from the collapsed tube in the upstream and downstream directions to prevent any influence of the imposed boundary conditions on the region of interest. This results in a division of the computational domain into three regions (see Fig.~\ref{fig:fluidDomain}): inlet, sound generation and propagation region, and acoustic suppression zone (ASZ) outlet. The sound generation and propagation region defines the region of interest. The inlet and outlet are treated as free stream boundaries, which allow to impose the flow velocity ($10$~m/s) and static temperature ($308$~K). One advantage of free stream boundaries is that they are perfect non-reflective boundaries for plane waves. As discussed in the following, the geometry and boundary conditions of the problem allow only plane waves to propagate in the acoustic frequency range. The flow pressure fluctuations, characterized by a much smaller length scale, are dissipated by the mesh stretching in the ASZ and, therefore, will not be reflected. All the other walls are treated with no-slip boundary conditions.

The simulation solver and mesh grid are tuned to perform an unsteady compressible flow Direct Noise Calculation (DNC), which allows the whole aeroacoustics characterization of the system to be performed within one simulation. Both the unsteady flow, which causes the noise generation, and the resulting radiated sound waves are solved simultaneously. This approach imposes precise requirements on the mesh grid, which is built in three steps:

\begin{enumerate}
    \item A coarse unstructured mesh of polyhedral elements is defined to discretize the system. It allows to obtain a well-converged preliminary solution based on the Reynolds Averaged Navier-Stokes (RANS) approach that will be used as the initial condition for the following unsteady simulation.
    \item Since the time evolution of the fluid features is resolved via a segregated Large Eddy Simulation (LES) model with a WALE sub-grid scale model, a first mesh refinement is implemented in the region of interest to resolve the Taylor micro-scale. Moreover, 10 prism layers are implemented on the walls to ensure well-resolved boundary layers.
    \item To ensure that all the waves in the acoustic frequency range can propagate in the domain, the mesh is further refined to ensure that at least $ 20$~grid cells resolve the shortest acoustic wavelength.
\end{enumerate}

\noindent
Outside the region of interest, the mesh is coarser and is stretched in the ASZ. In this way, the smaller-scale aerodynamic fluctuations are numerically dissipated away from the region of interest and cannot reflect into the domain.

The choice of the time step can also influence the numerical dissipation of acoustic waves. An implicit second-order time marching scheme is implemented in the model. A minimum of $15$~time steps per period of the highest frequency sound wave are recommended for a DNC simulation. For this simulation, the time step has been set to $\Delta t = 3.1\times10^{-6}$~s, corresponding to about $16$~time steps for a $20000$~Hz acoustic wave.

The presented fluid model has been implemented in the commercial software Star-CCM+ by Siemens. The fluid flow LES based solver has been previously validated in similar geometries by comparing the pressure predictions to corresponding experimental mid-line pressure data~\cite{schickhofer2020analysis}.

Velocity and pressure fields are extracted to calculate time-averaged and fluctuating components on relevant cross-sections for the following post-processing. The full-time history of pressure and velocity fields on the domain surfaces are stored at each time step to compute the surface Fourier transform. Several monitoring points for both velocity and pressure fields are placed in the domain.

\section{Sound generation mechanisms}
\label{sec:soundGen}
\begin{figure}[h!]
    \centering
    \includegraphics[width=\textwidth]{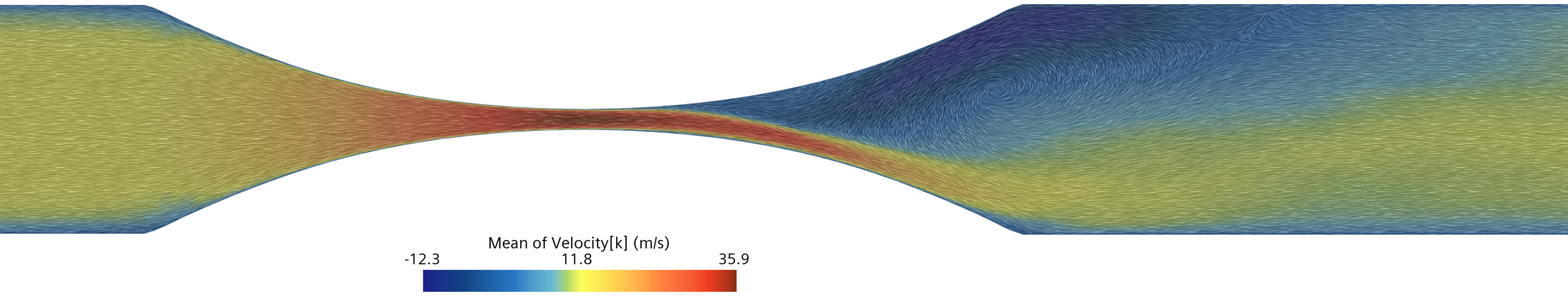}
    \includegraphics[width=\textwidth]{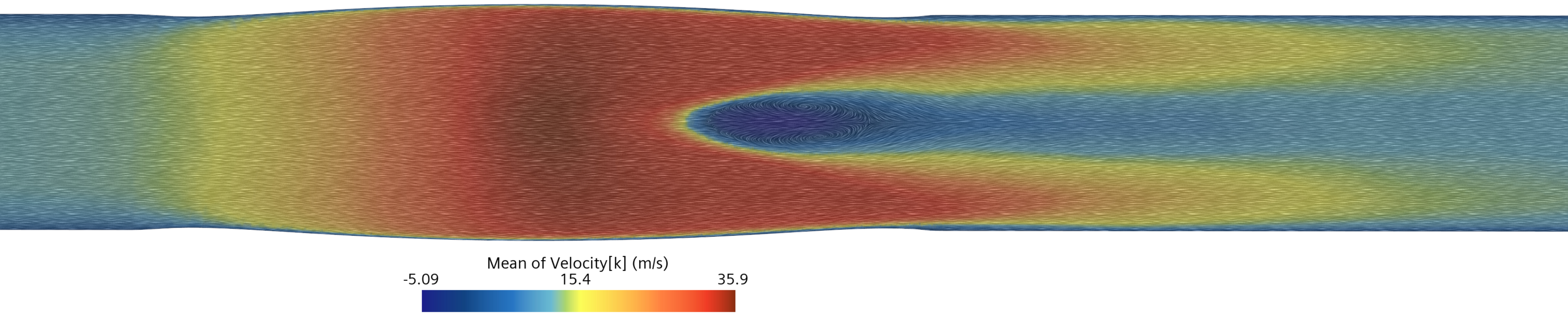}
    \caption{(Color online) Mean axial velocity in the two main longitudinal cross-sections of the domain in post-buckling configuration. The flow is characterized by the presence of Coand\u{a} effect. The arrows are obtained via line integral convolution of the velocity field and represent the mean velocity.}
    \label{fig:PostBMean}
\end{figure}

\begin{figure}[h!]
    \centering
    \includegraphics[width=\textwidth]{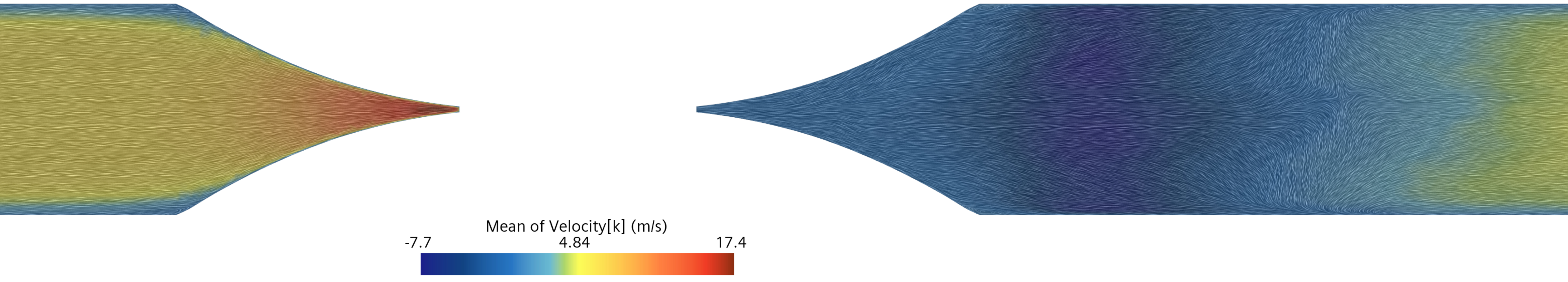}
    \includegraphics[width=\textwidth]{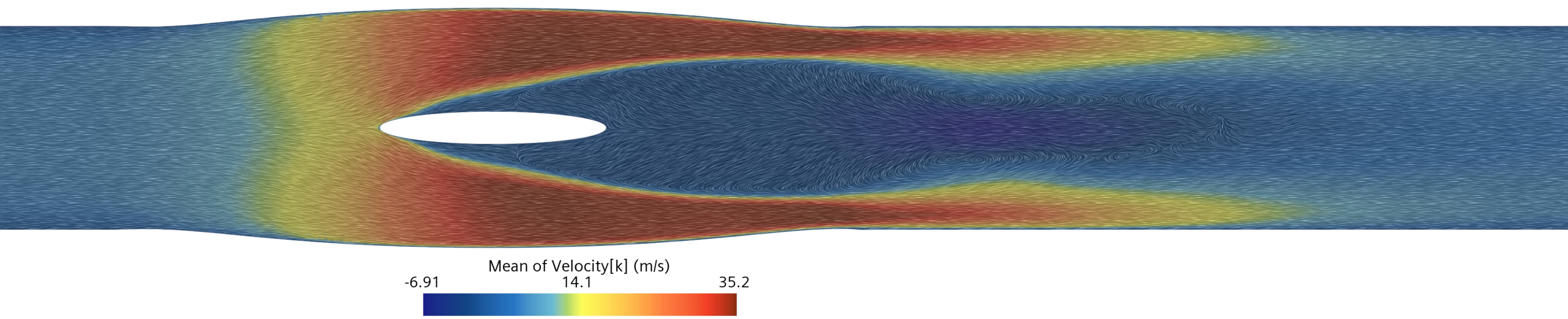}
    \caption{(Color online) Mean axial velocity in the two main longitudinal cross-sections of the domain in post-contact configuration. The empty regions correspond to the location of the contact of the lumen. The arrows are obtained via line integral convolution of the velocity field and represent the mean velocity.}
    \label{fig:PostCMean}
\end{figure}

The maximum velocity magnitude of the airflow under the three different collapse states of the tube is of the order of $40$~m/s, corresponding to a low Mach number regime ($\text{Ma}\simeq0.12$). Under such conditions, sound is generated mainly by the unsteady pressure loads caused by the flow vortical structures washing the walls of the domain. Therefore, analyzing such interaction in the three geometries under investigation is interesting. To this end, the distribution of the mean axial component of the velocity is analyzed on two orthogonal longitudinal cross sections (see Figs.~\ref{fig:PostBMean}-\ref{fig:PostCMean}), corresponding to the \textit{lateral view} (YZ-plane) and the \textit{top view} (XZ-plane). 

The Poiseuille-like flow in the pre-buckling configuration of the tube does not show any interesting flow feature. It is irrelevant for sound generation as it shows negligible time-dependent interactions with the walls. Consequently, it will not be further discussed in this section. The flow in the post-buckling configuration (Fig.~\ref{fig:PostBMean}) is characterized by a clear Coand\u{a} effect. This effect describes the tendency of a jet flowing from an orifice to attach to one side of the domain, developing a recirculation region on the other side. As shown in the YZ-plane cross-section (\textit{lateral view}, upper panel in Fig.~\ref{fig:PostBMean}), the buckled configuration acts on the fluid like a sudden contraction-sudden expansion zone. The resulting jet stays attached to the lower part of the domain (due to the Coand\u{a} effect), producing a large recirculation area in the YZ-plane. As discussed in Sec.~\ref{sec:onsetwaves}, the Coand\u{a} effect is responsible for the onset of acoustic waves in the domain. The cross-section in the XZ-plane (\textit{top view}, lower panel in Fig.~\ref{fig:PostBMean}) shows the symmetric behavior of the airflow through the two lobes of the buckled configuration.

In the post-contact configuration of the tube, the lumen is partially closed. Such a critical change in the geometry corresponds to a clear change in the fluid behavior. Due to the contact of the internal walls of the tube, the Coand\u{a} changes its direction (see the upper panel Fig.~\ref{fig:PostCMean}, \textit{lateral view}). As shown in the cross-section in the XZ-plane (lower panel in Fig.~\ref{fig:PostCMean}, \textit{top view}), the contact region acts like an adverse pressure gradient region, causing the separation of the boundary layers from the wall. Consequently, a large recirculation area is produced, constraining the flow towards the lateral walls of the domain and, in turn, generating sound.

\begin{figure}[h!]
    \centering
    \includegraphics[width=\textwidth]{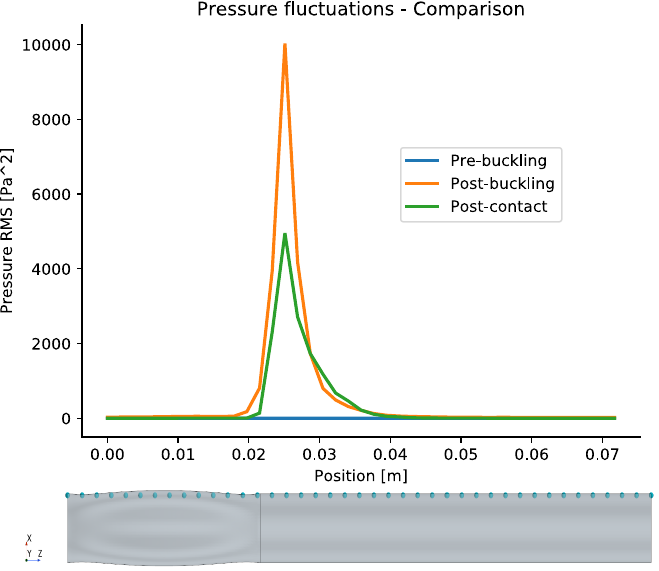}
    \caption{Top panel: Pressure RMS comparison between the three different geometries. The post-buckling configuration shows the largest pressure fluctuations on the wall. Bottom panel: position of the probe line where the pressure RMS has been calculated.}
    \label{fig:pRMS}
\end{figure}

Interestingly, the topology change from the post-buckling to the post-contact configuration of the collapsible tube induces a rotation of $90^\circ$ of the main recirculation area. The periodic flow separations in the post-buckling and post-contact configurations cause an oscillating force on the fluid in the perpendicular direction to the flow. Such unsteady force is the primary sound generation mechanism in a collapsible tube under physiological conditions. It is possible to estimate the unsteady pressure loads developed on the walls in the three collapse configurations by analysing the pressure root means square (RMS) on a probe line located on one boundary of the domain (see the bottom panel in Fig.~\ref{fig:pRMS}). Remarkably, the post buckling configuration shows much higher values for the pressure fluctuations on the wall, suggesting that this configuration generates higher sound power level with respect to the post-contact configuration. To prove this point, an in-depth sound power analysis is presented in the next section.

\section{Sound power analysis}  
\label{sec:sourceChar}
The analysis of the velocity field discussed in the previous section shows that the flow characteristics and the corresponding sound generation mechanisms depend dramatically on the collapse state of the tube. The sound power level can be computed by processing the results of the unsteady compressible CFD simulations, analysis carried out for all three different collapsed geometries under analysis. The analysis is based on the employment of the source field Lighthill's acoustic analogy. This entails that any conversion of acoustic energy into flow kinetic energy is neglected, as well as the sound field interference on the fluid flow. It is also assumed that the effects of the fluid motion on the sound propagation in the source region are not considered. 

The present section contains an analysis of the sound power level generated by the fluid flow in the near-fields under the different collapse phases. The investigated domain can be interpreted as the acoustic near-field. Indeed, as will be shown in the next section, the wavelength corresponding to the acoustic pressure fluctuations is of the same order as the longitudinal dimension of the domain. The following section will present the study of the subsequent propagation of acoustic waves in the domain. It will exploit the full power of the DNC simulations, which describe the coupling between the sound and the flow fields.

Lighthill's equation is a reformulation of the compressible Navier-Stokes equations in the form of an inhomogeneous wave equation expressed in terms of acoustic pressure fluctuation:
\begin{equation}
\label{eq:lighthill}
\begin{split}
    &\frac{1}{c_0^2}\frac{\partial^2 p'}{\partial t^2}-\frac{\partial^2 p'}{\partial x_i \partial x_i}=\\
    &\frac{\partial}{\partial t}\left(\frac{1}{c_0^2}\frac{\partial}{\partial t}(p'-c_0^2\rho')\right)-\frac{\partial g_i}{\partial x_i}+\frac{\partial^2}{\partial x_i \partial x_j}(\rho u_i u_j-\tau_{ij})\,.
\end{split}
\end{equation}
Here, the prime variables $\rho'$ and $p'$ are acoustic fluctuations of the density and pressure mean-field, respectively, $c_0$ is the speed of sound, $g_i$ represents the forces acting on the fluid, and $\tau_{ij}$ is the deviatoric stress tensor. The source term in Eq.~\ref{eq:lighthill} can be decomposed into the following terms:
\begin{equation}
\begin{split}
    &s_m=\frac{\partial}{\partial t}\left(\frac{1}{c_0^2}\frac{\partial}{\partial t}(p'-c_0^2\rho')\right)\\
    &s_d=-\frac{\partial g_i}{\partial x_i}\\
    &s_q=\frac{\partial^2}{\partial x_i \partial x_j}(\rho u_i u_j-\tau_{ij}).
\end{split}
\end{equation}
These terms correspond to the $n=0,1,2$ terms of a multi-pole expansion of order $2^n$, i.e., monopole, dipole, and quadrupole source terms. The monopole term $s_m$ describes sound source mechanisms due to deviation from adiabatic assumption, and it vanishes in the present analysis. The low Mach number regime of the system implies that the contribution of the quadrupole forces (which describe the sound produced by turbulence) is negligible compared to the dipole term. Indeed, the ratio of sound power generated by a dipole source to the one generated by a quadrupole source scales as Ma$^{-2}$, which implies that the contribution due to quadrupole sources to the total sound power level is in this case of about $2\%$. Consequently, the main contribution to the sound power generated by the flow in a collapsible tube under physiological conditions comes from the dipole source $s_d$. 

\begin{figure}[h!]
    \centering
    \includegraphics[width=\textwidth]{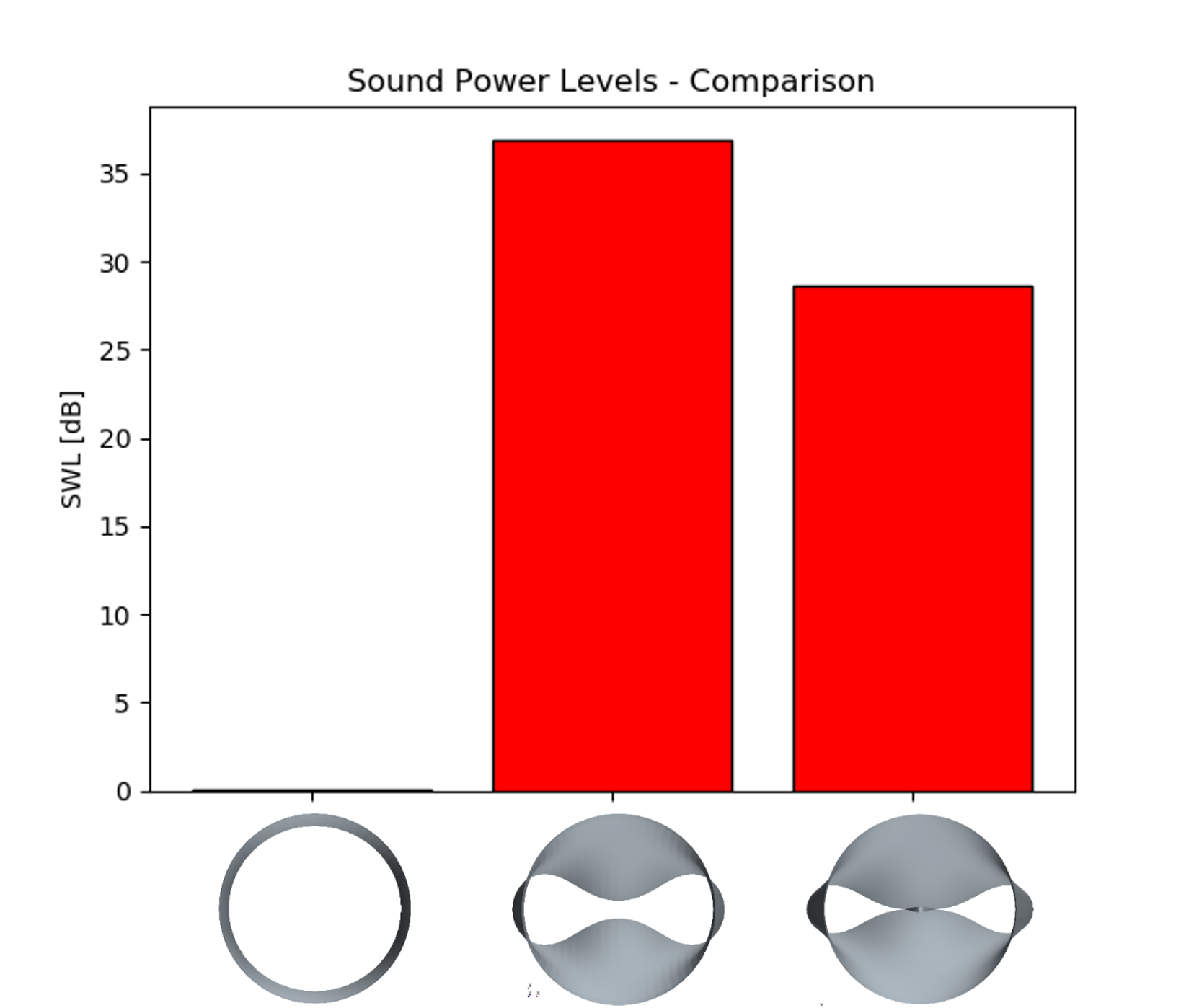}
    \caption{(Color online) Comparison of sound power level (SWL) defined as $10\log(W_d)$, where $W_d$ is defined in Eq.~\ref{eq:dipole} in the three collapse states under analysis. The value for the pre-buckling is smaller than the reference and has been set to zero.}
    \label{fig:SWL}
\end{figure}

The dipole source term describes the sound produced by the fluctuating forces due to flow separation from a rigid wall. As anticipated in the previous section, the periodic character of the developed flow structures in the generated shear-layer, as well as the flapping nature of the developed shear-layer washing the wall, causes an oscillating force on the walls in the perpendicular direction with respect to the flow. Another consequence of the low Mach number regime is that the collapsible tube acts as a compact acoustic source. Indeed, if $T$ is the typical temporal scale for sound production and $U$ is the typical mean flow velocity, the typical acoustic frequency is $f=1/T\sim U/D$, where $D$ is the diameter of the tube. The corresponding wavelength is $\lambda=c_0/f\sim D/\text{Ma}$, which implies that the source region is much smaller than the typical wavelength for low Mach numbers. Under these assumptions, the sound power radiated by a dipole source can be written as~\cite{aabom2006introduction}

\begin{equation}
\label{eq:dipole}
    W_d=\frac{\overline{\Dot{F}^2}}{12\pi\rho_0 c^3_0}
\end{equation}

\noindent
where the dot represents the time derivative, and $F^2=|\textbf{F}|^2$, i.e., the (squared) magnitude of the total force acting on the fluid, $F_i=\int_V g_i dV$, where $V$ is the volume of the fluid domain. It is possible to compute this term by extracting the total force acting on the fluid from the CFD simulations for the pre-buckling, post-buckling, and post-contact configurations. The resulting sound power levels are plotted in Fig.~\ref{fig:SWL}.

Interestingly, the maximum of the sound power level corresponds to the post-buckling configuration. The explanation is related to the stronger interaction between the flow and the walls in the post-buckling configuration. Indeed, as discussed in the previous section, the numerical model predicts larger root mean square pressure, with respect to the post-contact configuration, in the region located downstream of the minimum cross-sectional area where the developed shear-layers are interacting with the walls (see Fig.~\ref{fig:pRMS}). Larger pressure fluctuations result in higher values of $\overline{\Dot{F}^2}$ in Eq.~\ref{eq:dipole} and, in turn, in larger sound power levels. Such observation implies an \textit{optimal} collapsed tube configuration, which yields maximum sound power between the post-buckling and the post-contact phase. This suggests the existence of an \textit{acoustic tube law}, relating the intramural pressure (or the cross-section area) to the sound power radiated by the collapsible tube. The derivation of the acoustic tube law will be the object of future work, in which the system will be studied in an FSI setting with a continuous collapse. 
It is important to remark that the value computed using Eq.~\ref{eq:dipole} represents the total acoustic energy generated by the flow. However, it does not specify how such an energy will be radiated. Partially, it will generate acoustic waves, partially it will excite the domain walls, resulting in aeroelastic perturbations, and partially will be dissipated. The far-field propagation of the sound, both within the lumen and outside, will be investigated in future work.

\section{Onset of acoustic waves}
\label{sec:onsetwaves}

The sound generation mechanisms described in Sec.~\ref{sec:soundGen} induce pressure fluctuations in the collapsible tube. The DNC setup implemented for the numerical simulations allows to analyse at the same time the aerodynamic and acoustic components of such pressure fluctuations. This section analyzes the spectral properties of the pressure field in the tube under different collapse states. Moreover, the onset of acoustic waves is observed.

\begin{figure}[h!]
    \centering
    \includegraphics[width=\textwidth]{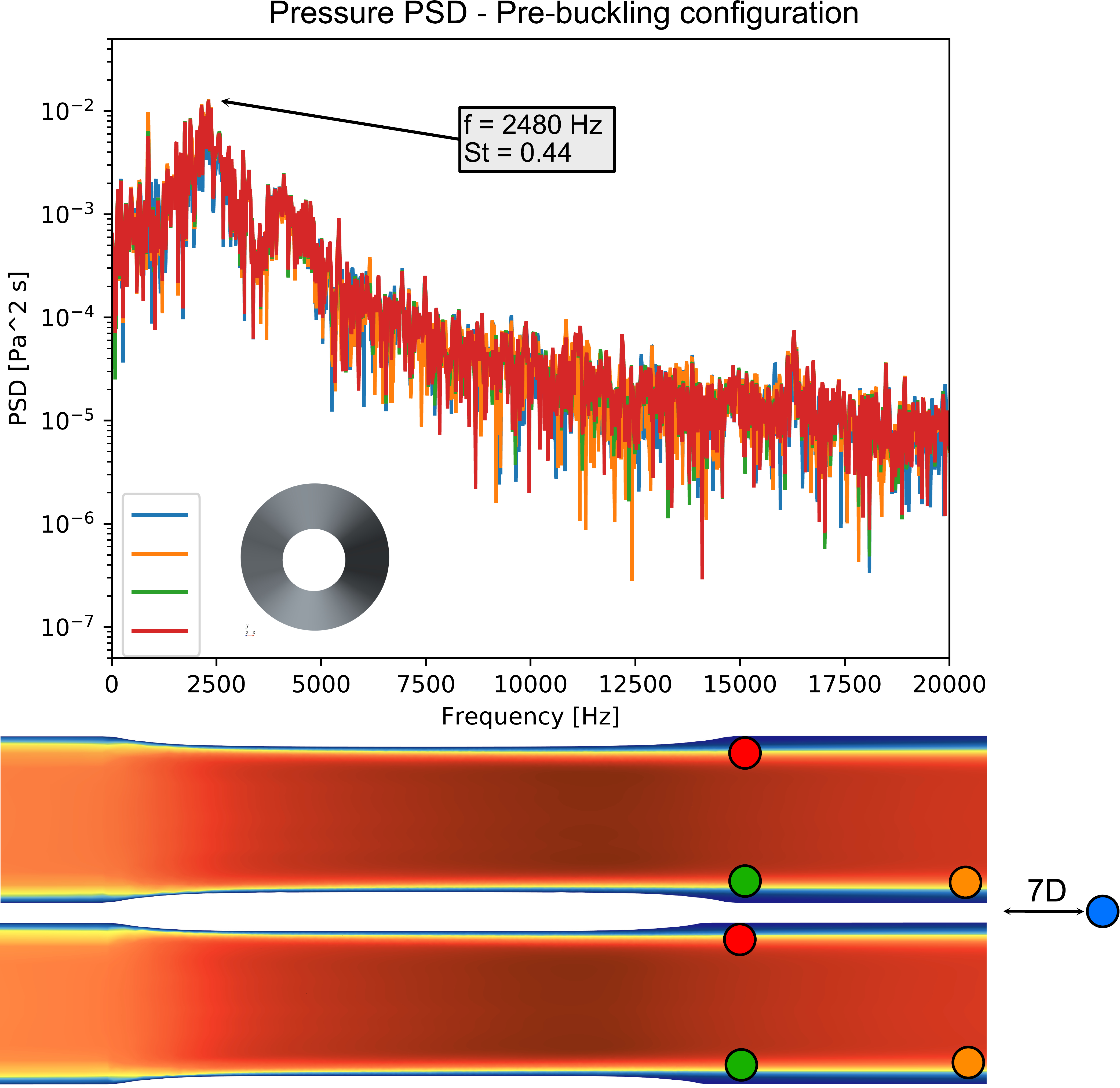}
    \caption{(Color online) Upper panel: Pressure PSD magnitude for the pre-buckling configuration. Lower panel: instantaneous axial velocity and probe positions in the domain. The probes' colors correspond to the colors in the PSD plots.}
    \label{fig:PSD_PreB}
\end{figure}

\begin{figure}[h!]
    \centering
    \includegraphics[width=\textwidth]{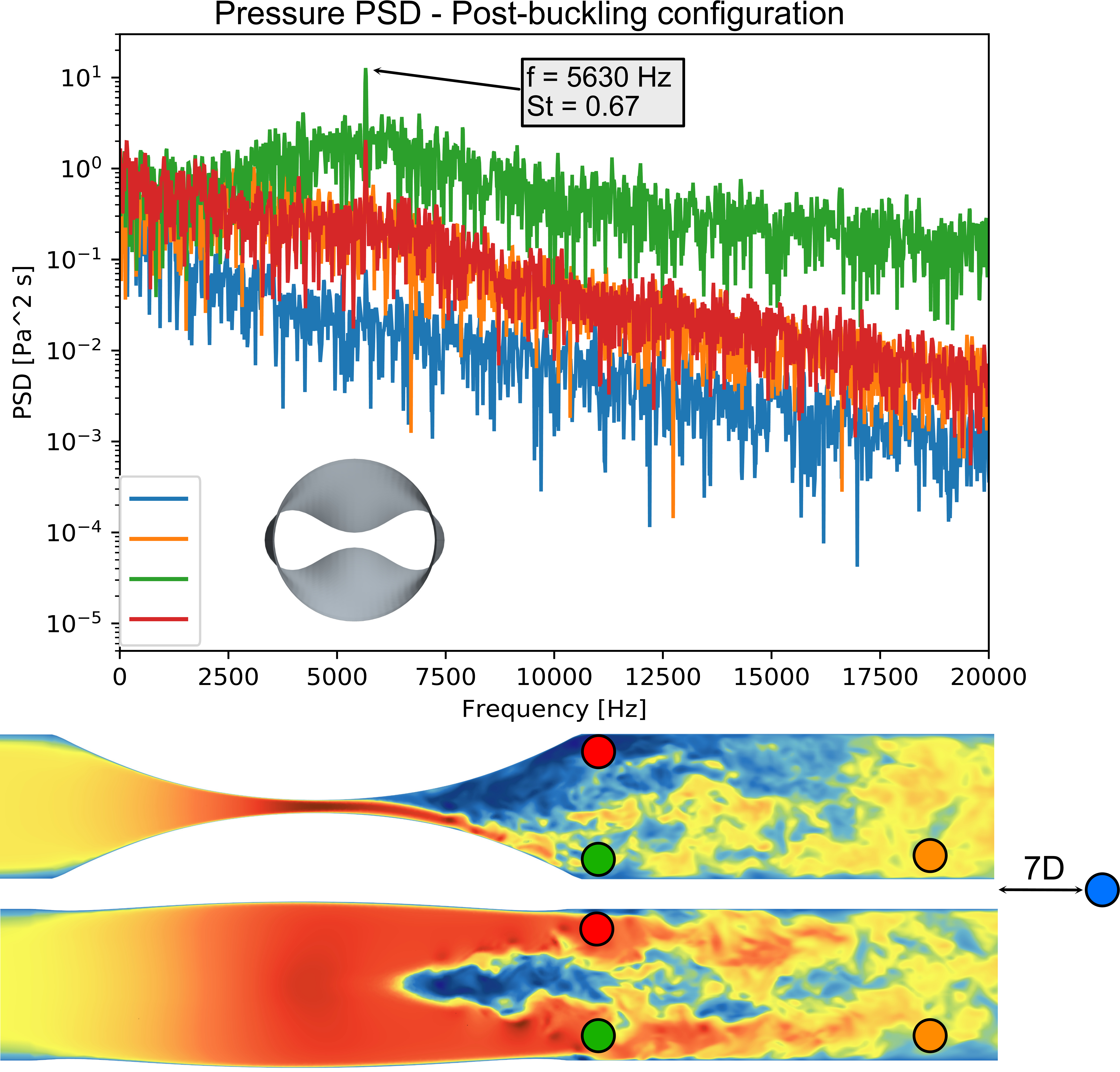}
    \caption{(Color online) Upper panel: Pressure PSD magnitude for the post-buckling configuration. Lower panel: instantaneous axial velocity and probe positions in the domain. The probes' colors correspond to the colors in the PSD plots.}
    \label{fig:PSD_PostB}
\end{figure}

\begin{figure}[h!]
    \centering
    \includegraphics[width=\textwidth]{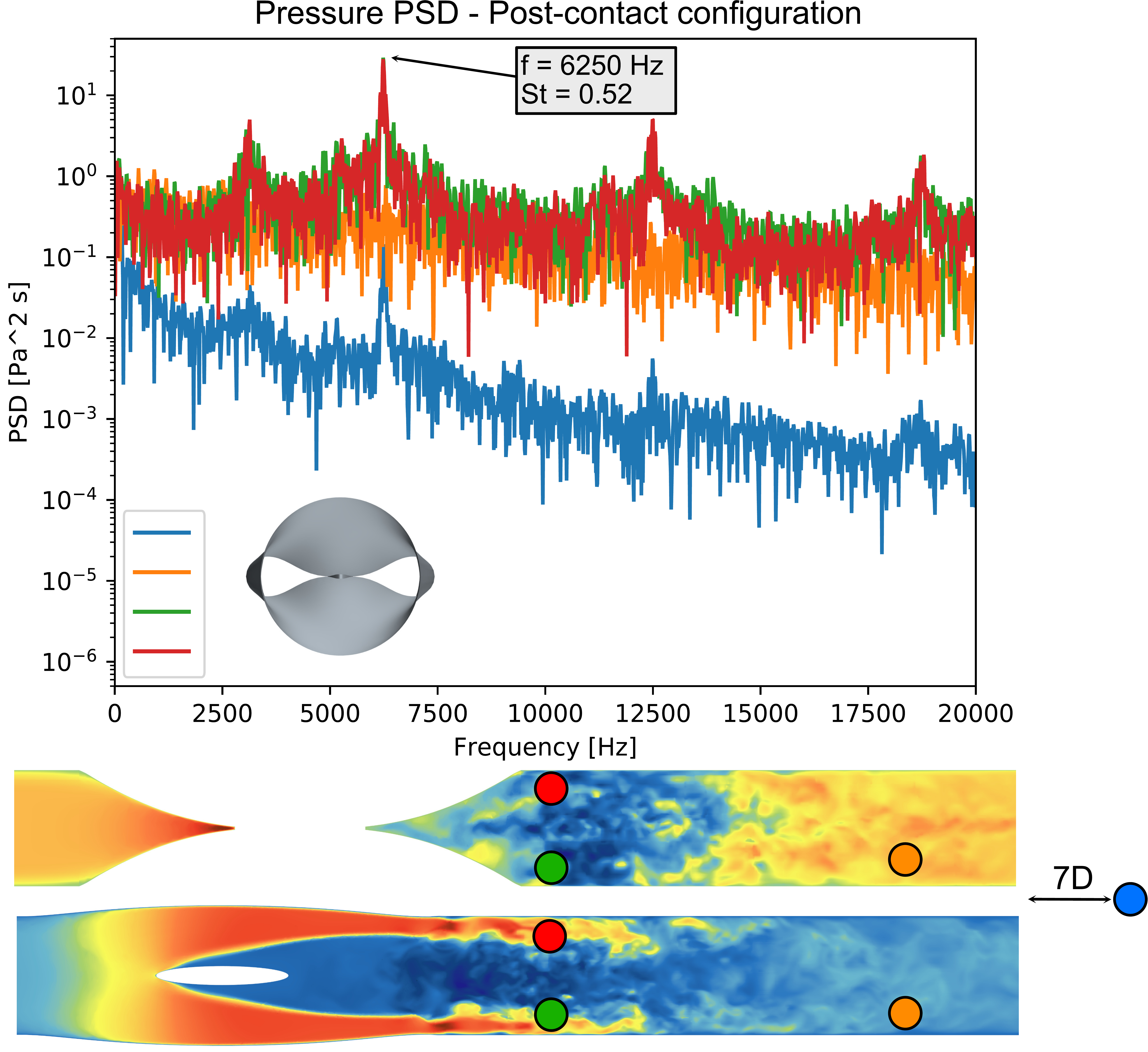}
    \caption{(Color online) Upper panel: Pressure PSD magnitude for the post-contact configuration. Lower panel: instantaneous axial velocity and probe positions in the domain. The probes' colors correspond to the colors in the PSD plots.}
    \label{fig:PSD_PostC}
\end{figure}

Probing points for the pressure field are positioned in sensible positions in the domain to capture the fluctuations associated with the vortical structures developed in the shear-layer (see Figs.~\ref{fig:PSD_PreB}-\ref{fig:PSD_PostC}). From the time history of the pressure signals, it is possible to obtain the corresponding power spectral density (PSD) for the three different collapse states of the tube.

A first remark is that the PSD for the pre-buckling configuration has negligible amplitudes compared to the more collapsed states of the tube, as expected from the negligible unsteadiness of the flow behavior in this configuration. The second observation is that all the PSD plots show narrow-band peaks. The corresponding frequencies are related to the small-scale fluctuations in the shear layers associated with the separation mechanisms described in Sec.~\ref{sec:soundGen}. The shear layer fluctuations are visible in correspondence of the probe points in the axial velocity snapshot plots in Fig.~\ref{fig:PSD_PostB} and~\ref{fig:PSD_PostC}. 

The PSD for the post-buckling configuration shows clear asymmetric amplitudes at the peak frequency $f = 5630$~Hz for the two near-field probes, which is explained by the asymmetric flow induced by the Coand\u{a} effect. Such asymmetry is not detected in the PSD levels for the post-contact configuration due to the disappearance of the Coand\u{a} effect, which induces a more symmetric flow behavior. It is interesting to notice that although the power density at the peak frequency in the post-contact configuration ($f = 6250$~Hz) is higher than the corresponding one in the post-buckling configuration, the analysis in Sec.~\ref{sec:sourceChar} shows that the overall energy of the pressure fluctuation in the post-buckling configuration is larger. 

The value of the peak frequencies in the post-buckling and post-contact configuration can be compared by means of the corresponding dimensionless frequency (Strouhal number) defined as
\begin{equation}
\label{eq:nondim_freq}
    \Tilde{f}=\frac{fD'}{U}\,,
\end{equation}
where $D'=\sqrt{2A_{min}/\pi}$ is the effective diameter, $A_{min}$ is the area of the central cross-section of the tube, and $U$ is the bulk flow velocity ($U=22$~m/s). The values of Strouhal number for the post-buckling and post-contact cases are $\Tilde{f}_{PB}=0.68$ and $\Tilde{f}_{PC}=0.52$, respectively. This difference is likely attributed to the different flow mechanisms related to the post-buckling and post-contact collapse state. Indeed, the flow behavior changes completely as the tube transitions from the post-buckling to the post-contact configuration. 

\begin{figure*}[h!]
    \centering
    \includegraphics[width=0.93\textwidth]{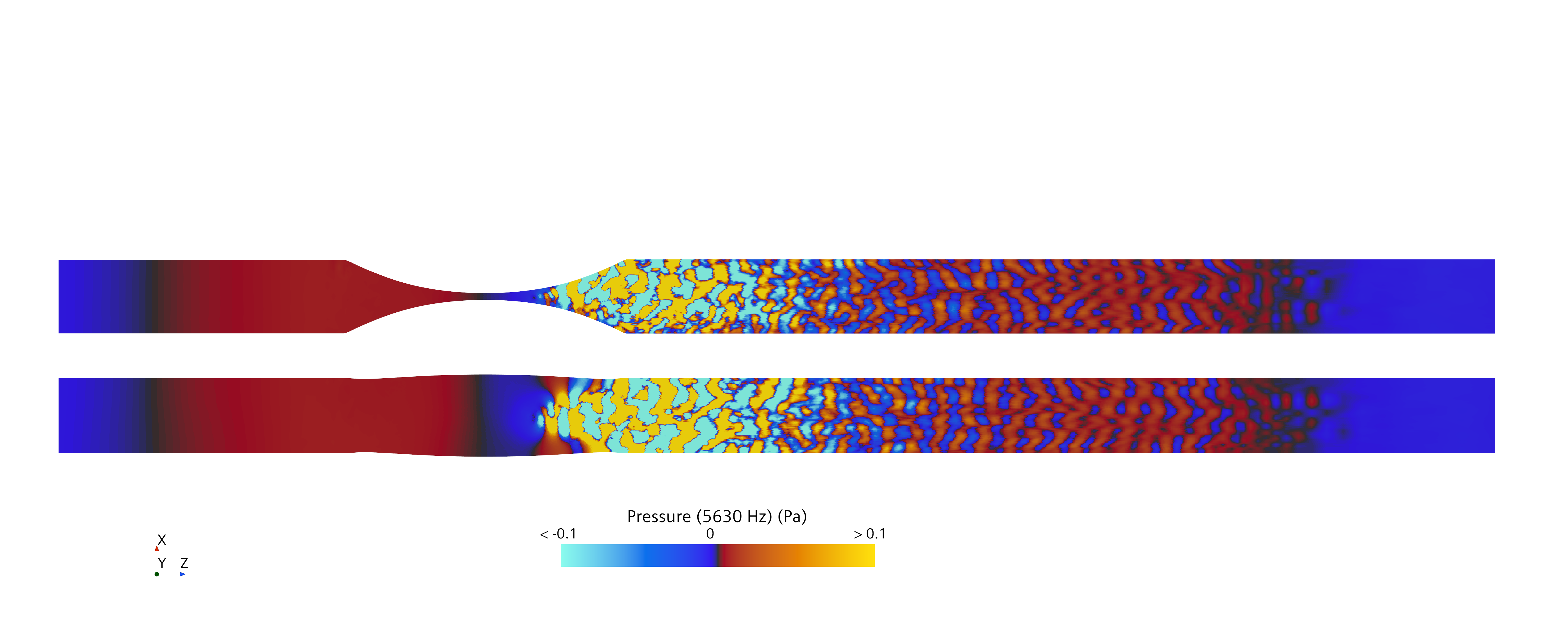}
    \caption{(Color online) Snapshot of the pressure fluctuations in the post-buckling configuration at $f=f_{PC}=5630$~Hz. The larger wavelength fluctuations propagate at the speed of sound.}
    \label{fig:pp_postB}
\end{figure*}

\begin{figure*}[h!]
    \centering
    \includegraphics[width=\textwidth]{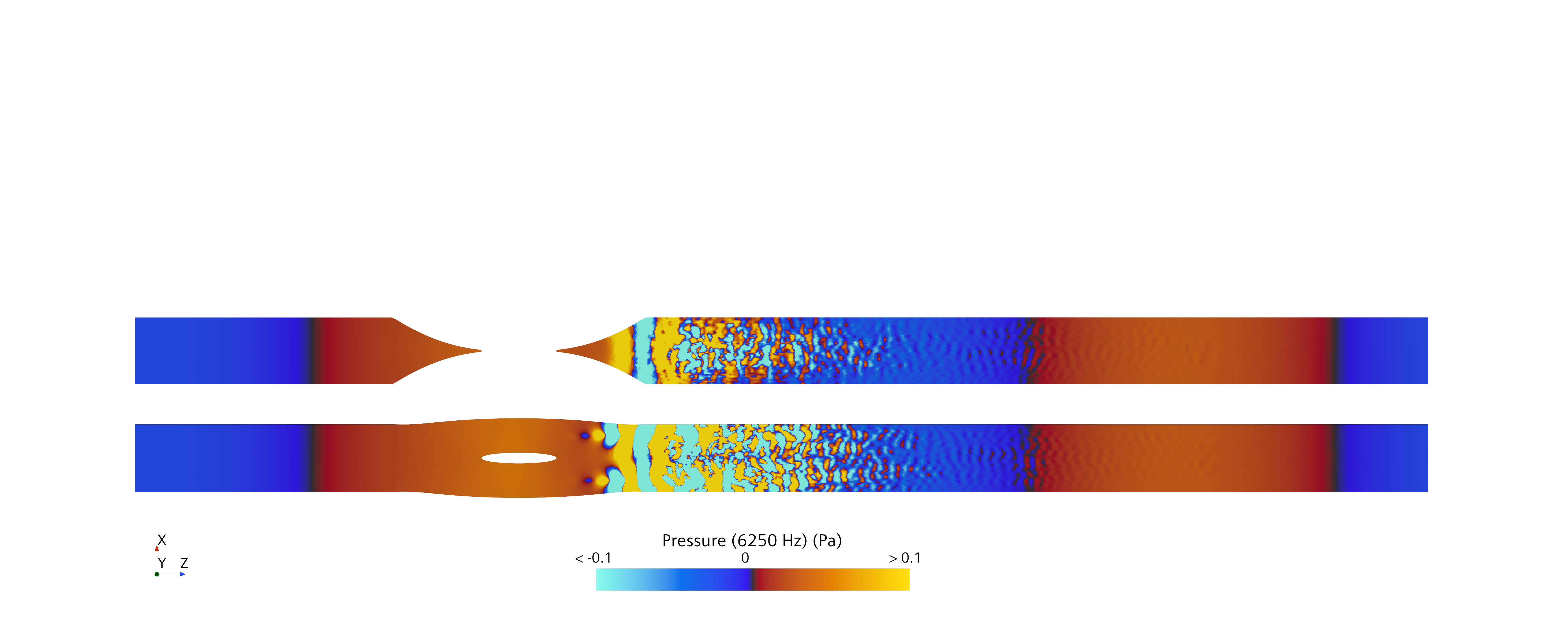}
    \caption{(Color online) Snapshot of the pressure fluctuations in the post-contact configuration at $f=f_{PC}=6250$~Hz. The larger wavelength fluctuations propagate at the speed of sound.}
    \label{fig:pp_postC}
\end{figure*}

To study the onset of acoustic waves within the domain induced by the flow features presented in Sec.~\ref{sec:soundGen}, it is possible to compute the Fourier transform of the pressure on the two main longitudinal cross-sections. The computation is performed using a built-in implementation in Star-CCM+. The pressure data is windowed with a Hann function, and the number of analysis blocks is $6$ with an overlap factor of $0.25$. Once the Fourier transform is obtained, it is possible to compute the corresponding real and imaginary parts evaluated at the two peak frequencies for the post-buckling and post-contact configurations, $f_{PB}=5630$~Hz and $f_{PC}=6250$~Hz, respectively. These can be used as the coefficients of the components of the Fourier series of the pressure time evolution $p(t)$ at the frequency $f_{PB}$ and $f_{PC}$. In formulae:

\begin{equation}
\label{eq:fourierseries}
    p(t) = \text{Re}\left[\Hat{p}(f_i)\right]\cos\left(\frac{2\pi t}{T}\right)+\text{Im}\left[\Hat{p}(f_i)\right]\sin\left(\frac{2\pi t}{T}\right)\,,
\end{equation}

\noindent
where $\Hat{p}(f_i)$ is the Fourier transform of the pressure signal evaluated at the frequency $f_i$, where $i=\{PB,PC\}$. The corresponding pressure fluctuations at specific peak frequencies on the two main longitudinal cross-sections is displayed in Fig.~\ref{fig:pp_postB} and in Fig.~\ref{fig:pp_postC}, for the post-buckling ($f_i=f_{PB}$) and the post-contact ($f_i=f_{PC}$) configurations, respectively.

\noindent
Both configurations show aerodynamic and acoustic pressure fluctuations, indicating that acoustic waves are generated at the frequency corresponding to the flow shear-layer fluctuations. The short wavelength pressure features downstream of the constriction have a clear aerodynamic origin. Indeed, since Eq.~\ref{eq:fourierseries} shows the time evolution of one frequency component of the signal, it is possible to estimate the phase velocity $c_f$, as $\lambda f=c_f$, where $\lambda$ is the wavelength of the feature. At the same time, it is possible to identify pressure fluctuations characterized by waves with larger wavelengths moving at the speed of sound. These waves are most noticeable upstream of the constriction, where the aerodynamic fluctuations are absent. Hence, it is possible to conclude that the fluid dynamics features described in Sec.~\ref{sec:soundGen}, characterized by frequencies $f_{PB}$ and $f_{PC}$, induce the onset of acoustic waves within the domain. 

It is possible to further determine the nature of these acoustic waves by analyzing the cut-off frequency $f_n^c$ of the system, i.e., the minimum frequency that allows the propagation of higher-order modes of order $n$ in the collapsible tube. The corresponding formula is~\cite{aabom2006introduction}
\begin{equation}
\label{eq:cutoff}
    f_n^c=\frac{c_0k^\perp_n}{2\pi}\sqrt{1-\text{Ma}^2}\,,
\end{equation}
where $k^\perp_n$ represents the $n$-th eigenvalues of the Helmholtz equation obtained by assuming a harmonic time-dependent solution in the homogeneous wave equation. For a circular duct, the first eigenvalue is $k_1^\perp=1.841/r$, where $r=0.003$~m is the radius of the non-deformed part of the collapsible tube, where the wave propagation happens. Consequently, the corresponding cut-off frequency is of the order of $30$~kHz, which implies that only acoustic plane waves can propagate in the domain for the frequency range considered in this study.

\section{Discussion and conclusions}
\label{sec:conclusions}
The analysis of the flow-induced sound generation mechanisms in a collapsible tube have been performed using CFD and CAA methods. The different flow features associated to three collapse states of the tube have been identified and related to the generation of acoustic power. The analysis of the pressure fluctuation on the walls of the tube and the corresponding sound power analysis have shown that the maximum of the sound power is produced in the post-buckling configuration. This suggest the existence of a non-trivial acoustic tube law, with maximum sound power level located in the range of pressures corresponding to the non-buckling configuration. Moreover, the compressible LES model has shown the onset of acoustic plane waves at the frequencies of to the shear-layer fluctuations. The results obtained in this analysis open to new perspective research questions.

The relation between the airflow peak frequencies indicated in Figs.~\ref{fig:PSD_PreB}-\ref{fig:PSD_PostC} and the frequency of possible self-excited oscillations of the airways' wall, in turn related to wheezing~\cite{grotberg1989flutter}, is not straightforward. Heil et al.~\cite{heil2010self} have shown that a collapsible tube in the post-buckling configuration shows self-excited oscillations for large enough Reynolds numbers. However, understanding the underlying physical mechanisms for the onset of such oscillations is currently an open question~\cite{heil2011fluid}. A possible research question is how the energy content of the flow pressure fluctuations due to the unsteady flow interacting with one side of the collapsible tube in the post-buckling configuration is coupled with the solid structure. In other words, it is relevant to determine the frequency response function of a collapsible tube, and how is it related to the collapsible state of the system. The role of the flow-induced acoustic waves on the onset of self-excited oscillations can be also investigated. The analysis of the flow and acoustic features presented in this work represents a first step in this direction, and it will be addressed in a future investigation involving a two-way coupled FSI numerical model.

Another interesting perspective is related to the analysis of the non-dimensional frequency associated to the peaks in the PSD plots~\ref{fig:PSD_PreB}-\ref{fig:PSD_PostC} and the physics of the fluid features in the post-buckling and post-contact configuration of the tube. A more in-depth analysis can be designed to prove this thesis. The strategy would be to simulate the fluid flow in several instances of post-buckling and post-contact configurations and to compute the corresponding Strouhal number. The thesis would be proven if the Strouhal number is consistently equal to $\Tilde{f}_{PB}$ for all the post-buckling simulations and to $\Tilde{f}_{PC}$ for all the post-contact simulations. This analysis will be the object of a future investigation.

\section*{Conflict of interest}
The authors declare no conflict of interest.

\section*{Data availability}
The simulation data can be provided under reasonable request.

\section*{Acknowledgments}
The project is financed by KTH Engineering Mechanics in the thematic area of Biomechanics, Health, and Biotechnology (BHB). Laudato is supported by the Swedish Research Council, grant No. 2022-03032. Zea is partly supported by the Swedish Research Council, grant No. 2020-04668. The authors acknowledge PRACE for awarding access to the Fenix Infrastructure resources at CINECA, which are partially funded by the European Union’s Horizon 2020 research and innovation programme through the ICEI project under grant agreement No. 800858. The simulations were partly run on the Swedish National Infrastructure for Computing (SNIC) resources at the PDC Centre for High Performance Computing (PDC-HPC).

\printbibliography


\end{document}